\RequirePackage{fix-cm}
\documentclass[smallextended]{svjour3}       
\smartqed  
\usepackage{graphicx}
\usepackage{amssymb}
\usepackage{mathptmx}

\usepackage{bm}
\newcommand{\eq}[1]{(\ref{#1})}
\newcommand{\Eq}[1]{Eq.~(\ref{#1})}
\newcommand{\Fig}[1]{Fig.~\ref{#1}}
\newcommand{\Ref}[1]{Ref.~\cite{#1}}
\def\beq{\begin{equation}} \def\eeq{\end{equation}}
\def\bea{\begin{eqnarray}} \def\eea{\end{eqnarray}}
\def\bse{\begin{subequations}} \def\ese{\end{subequations}}

   \def\vecr{{\bm r}} \def\vecv{{\bm v}}\def\vecD{{\bm \nabla}}
    
\def\||{\parallel}

\def\<{\left\langle} \def\>{\right\rangle}
\def\({\left(} \def\){\right)}
\def\[[{\left[} \def\]]{\right]}

\begin{document}

\title{On internal structure of smaller domains in domain coarsening dynamics of spontaneous $Z_2$-symmetry breaking in two dimensions
}


\author{Hiromitsu Takeuchi
}


\institute{H. Takeuchi \at
              Department of Physics, Osaka City University, 3-3-138 Sugimoto, Sumiyoshi-ku,\\ Osaka, 558-8585, Japan \\
              \email{hirotake@sci.osaka-cu.ac.jp}           
 }

\date{Received: date / Accepted: date}

\maketitle

\begin{abstract}
The internal structure of domains smaller than the characteristic size in domain coarsening dynamics of $Z_2$ symmetry breaking is evaluated theoretically for different phase ordering systems in two dimensions.
In the previous works on (non-) conserved Ising systems and binary superfluids,
 the statistical properties of smaller domains are analyzed by assuming that the contribution from its internal structure is negligible.
It is shown that this assumption is justified analytically with respect to the statistical quantities, such as domain area, domain-wall length and superfluid circulation, according to the empirical dynamic scaling law for the smaller domains.
\keywords{Spontaneous symmetry breaking \and phase ordering kinetics \and multi-component superfluids}
\end{abstract}

\section{Introduction}
 Spontaneous $Z_2$-symmetry breaking is one of the most fundamental problems in the field of statistical physics and is discussed universally in different physical systems from low temperature physics to cosmology and high energy physics \cite{1995Vilenkin,2000Bunkov,2002Onuki,2006Vachaspati}.
When a system undergoes the symmetry breaking by being quenched from the disordered phase into the ordered phase,
domain walls are nucleated as topological defects in the order parameter, represented with a real scalar field, by forming a complicated {\it texture} due to the Kibble-Zurek mechanism \cite{1976Kibble}\cite{1985Zurek}.

According to the dynamic scaling hypothesis of phase ordering kinetics \cite{1994Bray},
the coarsening development in the dynamic scaling regime after the defect nucleation is universally characterized by the dynamic exponent $z$, which determines the dependence of the mean distance $l$ between defects on time $t$ as $l(t)\propto t^{1/z}$.
In the dynamic scaling regime,
the spatial structure of the order parameter fields keeps its statistical similarity during the highly non-equilibrium development; the distributions of topological defects is statistically similar between different times after rescaling the length by $l(t)$.
This argument is applied to different systems of spontaneous symmetry breaking, {\it e.g.}, a coarsening development of vortex strings due to $U(1)$ symmetry breaking in three dimensions.

Recently, it is hypothesized that a coarsening system of $Z_2$-symmetry breaking in two dimensions generally possesses a hierarchy with the two scaling regimes in the domain-area distribution, namely, the macroscopic and microscopic regimes with domain sizes larger and smaller than $l$, respectively \cite{2018Takeuchi}.
The distribution in the former universally obeys a power law with the Fisher's exponent $\tau_{\rm mac}=\tau_{\rm F}=187/91\approx 2$ from the percolation theory \cite{1994Stauffer},
while that $\tau_{\rm mic}$ in the latter depend strongly on the coarsening system considered \cite{2018Takeuchi}\cite{2007Arenzon}\cite{2008Sicilia}\cite{2009Sicilia}\cite{2016Takeuchi}\cite{2017Bourges}.
Although it is expected that the value of $\tau_{\rm mic}$ is determined by the {\it microscopic} dynamics of domain walls in each system,
its physics and statistics are not well known.

In this work, we focus the internal structure of domains in the microscopic regime.
It is useful to clarify how the internal structure influences the statistical behavior of coarsening system.
The existence of domains embedded inside the circular domain can make it complicated to perform the scaling analysis.
In the literature \cite{2009Sicilia},
it was assumed that domains in the microscopic regime are independent and they do not have holes of the opposite domain within.
In the superfluid system \cite{2018Takeuchi},
 a dynamic scaling law for the distribution of superfluid circulation in circular domains in the microscopic regime has been derived by neglecting the existence of such embedded domains inside.
In these studies, the assumptions were made with no theoretical basis and
the argument were restricted to the case of each value of $\tau_{\rm mic}$.
Here, it is shown analytically that, for general cases of arbitrary value of the exponent $\tau_{\rm mic}$,
the influence of the internal structure is negligible with respect to domain area, it perimeter and superfluid circulation.

\section{The scaling behavior in the microscopic regime}

Considering a two dimensional system of domain coarsening,
 where a domain wall is a linear object existing between the two kinds of domains, $\uparrow$- and $\downarrow$-domains.
Consider the number of $\uparrow(\downarrow)$-domains, which have areas between $S$ and $S+dS$, divided by the system area $L^2$ at time $t$ as $\rho_{\uparrow(\downarrow)}(S,t)dS$.
Because of the statistical symmetry between the two domains, we have $\rho=\rho_{\uparrow}=\rho_{\downarrow}$ in the dynamic scaling regime of the coarsening development.
According to the dynamic scaling hypothesis,
the domain-area distribution is written by a universal dimensionless function, independent of time $t$, after rescaling the area $S$ by the characteristic area $S_l(t)=\pi l(t)^2$,
$
\tilde{\rho}(\tilde{S})=S_l^2\rho(S,t),
$ where we introduced the rescaled area $\tilde{S}=S/S_l$.
The distribution in the macroscopic regime $\tilde{S}\gg 1$ obeys the Fisher's power-law independent of the system as
 $\tilde{\rho}\sim \tilde{S}^{-\tau_{\rm F}}$ as was observed in the previous works \cite{2018Takeuchi}\cite{2007Arenzon}\cite{2008Sicilia}\cite{2009Sicilia}\cite{2016Takeuchi}\cite{2017Bourges}.
Contrastively,
in the microscopic regimes,
$
\tilde{S}_{\rm min}\equiv \frac{\pi l_{\rm min}^2}{S_l} \ll \tilde{S} \ll 1
$
with the thickness $l_{\rm min}$ of domain wall,
the exponent $\tau_{\rm mic}$ of the power law
\begin{eqnarray}\label{rho_mic}
\tilde{\rho}\sim \tilde{S}^{-\tau_{\rm mic}}
\end{eqnarray}
 differs dependent on coarsening systems as was mentioned in the introduction.
Here, $\tilde{S}_{\rm min}$ is the lower limit of the rescaled area $\tilde{S}$, under which
the domain is ill-defined.

According to \cite{2018Takeuchi}, the exponent $\tau_{\rm mic}$ has a theoretical upper limit for $\tilde{S}_{\rm min}\to 0$;
\begin{eqnarray}\label{tau_mic}
\tau_{\rm mic} < \frac{3}{2}.
\end{eqnarray}
This restriction for the exponent is derived from the normalization condition for the total length $R$ ($l\equiv L^2/R$) of domain walls,
$ 
\frac{R}{L^2}=\int  l_{\rm w} \rho  dS = \frac{1}{\pi l}\int  \tilde{l}_{\rm w} \tilde{\rho} d\tilde{S}
$ 
with the length $l_{\rm w}(S)\equiv \tilde{l}_{\rm w}l$ of domain walls that enclose a domain of area $S$.
It is empirically known that the domains are almost circular in the microscopic regime, and we have
\begin{eqnarray}\label{D_mic}
\tilde{l}_{\rm w} \sim \sqrt{\tilde{S}}.
\end{eqnarray}
Since the integral $\int  \tilde{l}_{\rm w} \tilde{\rho} d\tilde{S}$ must be on the order of unity,
the contribution from the microscopic regime ($\tilde{S}_{\rm min} \ll \tilde{S} \ll 1$) to the integral must be less than or on the order of unity.
Then, we have the restriction (\ref{tau_mic}) for the statistical limit $\tilde{S}_{\rm min}\to 0$.

\section{Contributions to the statistical quantities from the internal structure}
In this section, we evaluate the statistical quantities, domain area, domain wall length and superfluid circulation, contributed from the internal structure inside a circular domain in the microscopic regime under the restriction (\ref{tau_mic}).


\subsection{Domain area}
\begin{figure} 
\begin{center}
\includegraphics[width=1.0 \linewidth,keepaspectratio]{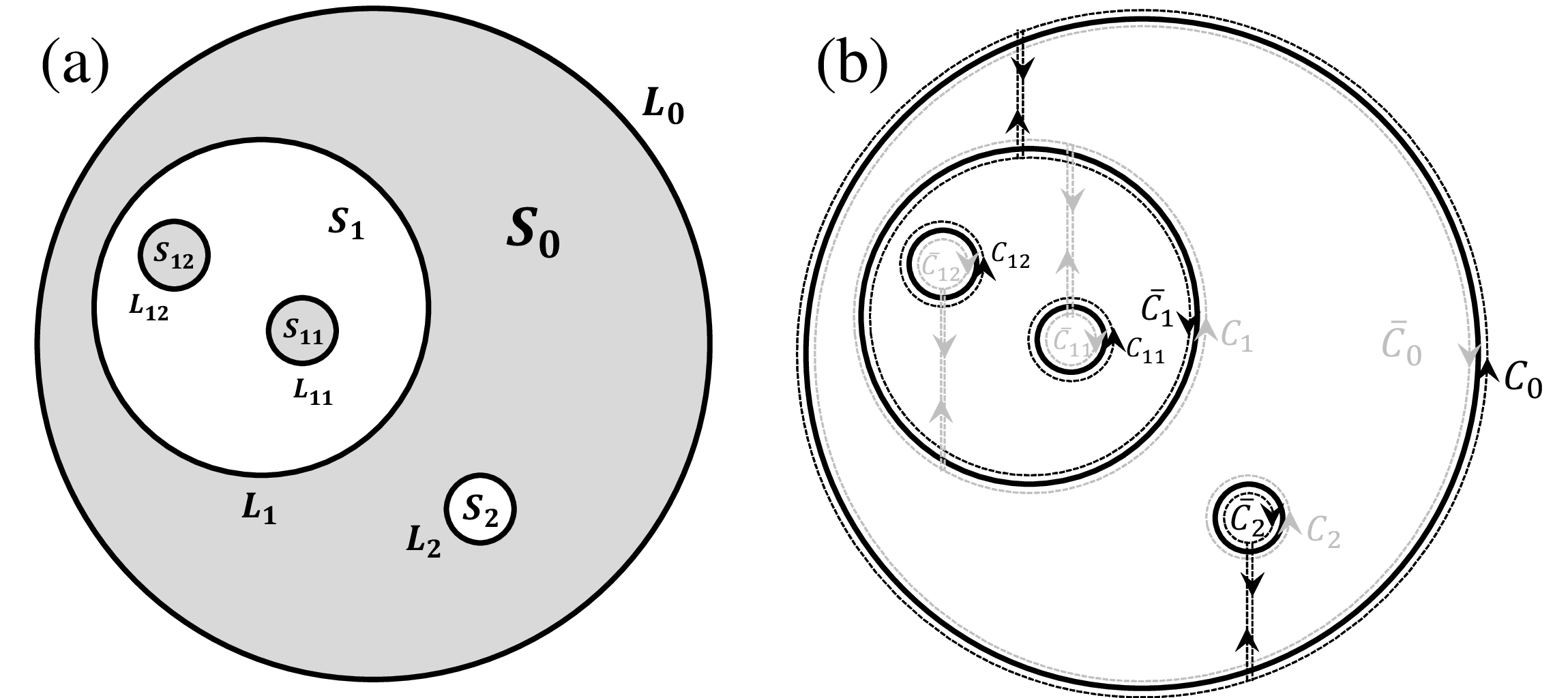}
\end{center}
\vspace{-5mm}
\caption{
Schematic diagram of a domain with area $S_0$ inside which several domains are embedded.
Light gray and white regions show $\downarrow$-domains and $\uparrow$-domains, respectively.
Domain walls are represented with bold black lines.
The circulation $\Gamma$ of the domain ($S_0$) is calculated along the line integral of the superfluid velocity $\vecv_\uparrow$ along the path $C_0+\sum_j\bar{C}_j$ (dashed gray and black lines).
In this schematic, we have $S_{j_\uparrow}=\{ S_1,S_2 \}$ and $S_{j_\downarrow}=\{ S_{11},S_{12} \}$.
}
\label{Fig_S_out}
\end{figure}

We consider a domain, immersed in the sea of a domain of a different kind.
Without loss of generality, we consider a circular domain wall surrounding a $\uparrow$-domain in the sea of a $\downarrow$-domain.
The $\uparrow$-domain of area $S_0(\ll S_l)$ may contain domains inside the circular, outer perimeter of length $L_0$, as is schematically illustrated in \Fig{Fig_S_out}~(a).

The total area $S_{\rm total}$, the area enclosed by the outer perimeter, is represented as
\begin{eqnarray}\label{Stotal}
S_{\rm total}=S_0+S_{\rm in},
\end{eqnarray}
where
$ 
S_{\rm in}=\sum_j \left\{ S_j +\sum_k\[[S_{jk}+\sum_l\(S_{jkl}+ \cdots \)  \]]   \right\}
$ 
 is the total area of all domains embedded in the domain $S_0$;
 $S_j~(j=1,2,3,...)$  is the area of a domain embedded in the domain $S_0$,
 $S_{jk}~(k=1,2,3,...)$ is the area of a domain embedded in the domain $S_j$, and such like.

The area of each embedded domain must be not much larger than $S_0$.
Otherwise, such a domain of $S_0$ forms an annulus, whose thickness $\Delta r$ is much smaller than $L_0$ with $S_0\approx \Delta r L_0$.
Such an {\it one-dimensional} object is conflict to the scaling law \eq{D_mic};
 we have $l_w\approx 2L_0\approx 2\frac{S_0}{\Delta r} =2\sqrt{S_0} \frac{\sqrt{S_0}}{\Delta r}=\sqrt{S_0}\sqrt{\frac{L_0}{\Delta r}}\gg  S_0^{1/2}$ with $\frac{L_0}{\Delta r}\gg 1$.
Then, the area $S_{\rm in}$ is statistically evaluated as
\begin{eqnarray}\label{S_in}
S_{\rm in} \lesssim 2S_{\rm total} \int_{S_{\rm min}}^{S_0} S \rho dS
=2S_{\rm total} \int_{\tilde{S}_{\rm min}}^{\tilde{S}_0} \tilde{S} \tilde{\rho} d\tilde{S}
\end{eqnarray}
with $\tilde{S}_0\equiv S_0/S_l$.
Here, the prefactor `$2$' comes from the fact that there are two kinds of domains inside.

By inserting \Eq{rho_mic} with \Eq{tau_mic} into \Eq{S_in}, we have
\begin{eqnarray}\label{area}
\frac{S_{\rm in}}{S_{\rm total}} \lesssim \tilde{S}_0^{2-\tau_{\rm mic}}-\tilde{S}_{\rm min}^{2-\tau_{\rm mic}} \ll 1.
\end{eqnarray}
Equation (\ref{area}) clearly shows that the contribution from the embedded domains is negligible, yielding
\begin{eqnarray}\label{Stotal_approx}
S_{\rm total} \approx S_0.
\end{eqnarray}

\subsection{Domain wall length}\label{Sec:length}
Next, we show that the contribution $L_{\rm in}$ to the domain-wall length from the embedded domains is also negligible statistically.
We write the total length of domain walls (see \Fig{Fig_S_out}(a) again) as
\begin{eqnarray}\label{Ltotal}
L_{\rm total}=L_0+L_{\rm in},
\end{eqnarray}
where
$ 
L_{\rm in}=\sum_j\left\{L_j +\sum_k\[[L_{jk}+\sum_l\(L_{jkl}+ \cdots \)  \]]   \right\}
$ 
 is the length contributed from all embedded domains.
The length $L_{\rm in}$ is calculated by summing lengths of all domain walls that surround embedded $\uparrow$-domains in \Fig{Fig_S_out}.
Thus, we have
\begin{eqnarray}\label{Lin}
L_{\rm in} \lesssim S_{\rm total} \int_{S_{\rm min}}^{S_0} l_{\rm w} \rho dS
=l\tilde{S}_{\rm total} \int_{\tilde{S}_{\rm min}}^{\tilde{S}_0} \tilde{l}_{\rm w} \tilde{\rho} d\tilde{S}
\end{eqnarray}
with $\tilde{S}_{\rm total}=S_{\rm total}/S_l$.
 From $S_{\rm total}\approx S_0\sim L_0^2$ with \Eq{Stotal_approx},
one obtains
\begin{eqnarray}\label{length}
\frac{L_{\rm in}}{L_0} \lesssim   \tilde{S}_0^{2-\tau_{\rm mic}}-\tilde{S}_0^{\frac{1}{2}}\tilde{S}_{\rm min}^{\frac{3}{2}-\tau_{\rm mic}} \ll 1,
\end{eqnarray}
which reduces to
\begin{eqnarray}\label{Ltotal_approx}
L_{\rm total}\approx L_0.
\end{eqnarray}
This result shows that the contribution of embedded domains is statistically negligible when one computes the wall length $l_{\rm w}$ of domains in the microscopic regime.

\subsection{Superfluid circulation}
 Finally, we evaluate the superfluid circulation around a domain in the microscopic regime of multi-component superfluid system, where vortex sheets exist along domain walls.
It is shown that the contribution from vortex sheets of the embedded domains is negligible.

The superfluid circulation $\Gamma_0$, namely the vorticity {\it charged} along the outer perimeter $L_0$, is represented by the line integral
\begin{eqnarray}\label{Gamma_out}
\Gamma_0
=\oint_{C_0}\vecv_{\uparrow}\cdot d\vecr+\oint_{\bar{C}_0}\vecv_{\downarrow}\cdot d\vecr
=\kappa n_{\rm v}(S_0)-\sum_j \Gamma_j
\end{eqnarray}
with the circulation quantum $\kappa$ and the superfluid velocity $\vecv_{\uparrow,\downarrow}$ in $\uparrow$- or $\downarrow$-domains [see \Fig{Fig_S_out}(b)].
The number $n_{\rm v}(S)=\kappa^{-1}\int_{S}  \vecD \times \vecv_{\sigma} dxdy~~(\sigma=\uparrow~{\rm or}~\downarrow)$ represents the total number of quantized vortices that exist {\it within} a domain of $S$.
Similarly, the vorticity $\Gamma_j$, charged along the perimeter $L_j$ in \Fig{Fig_S_out}, is rewritten as
 $\Gamma_j
=\oint_{C_j}\vecv_{\downarrow}\cdot d\vecr+\oint_{\bar{C}_j}\vecv_{\uparrow}\cdot d\vecr
=\kappa n_{\rm v}(S_j)-\sum_k \Gamma_{jk}$
with the vorticity $\Gamma_{jk}=\oint_{C_{jk}}\vecv_{\uparrow}\cdot d\vecr+\oint_{\bar{C}_{jk}}\vecv_{\downarrow}\cdot d\vecr$, charged along $L_{jk}$.
Definitely, \Eq{Gamma_out} reduces to
\begin{eqnarray}\label{Gamma_nv}
n_{\rm v}(S_0)=\frac{\Gamma_0}{\kappa}+N_\uparrow-N_\downarrow
\end{eqnarray}
with $N_{\sigma}=\sum_{j_{\sigma}}n_{\rm v}(S_{j_\sigma})$, where the sum $\sum_{j_{\sigma}}$ is taken over all embedded $\sigma$-domains.

To show that $N_{\sigma}$ is statistically negligible,
we apply the numerical result of dynamic scaling law based on the hydrodynamic theory \cite{2018Takeuchi},
\begin{eqnarray}\label{law_nv}
 | \tilde{n}_{\rm v}(\tilde{S}) | \equiv \tilde{S}_{\rm min}^{\frac{1}{4}}|n_{\rm v}(S,t)| \sim \tilde{S}^{\frac{1}{4}}~~~(\tilde{S}_{\rm min} \ll \tilde{S}\ll 1).
\end{eqnarray}
This law is applicable when almost domains in the microscopic regime have non-zero circulation in later stage of the coarsening development.
The contribution $N_\sigma$ to the circulation number $n_{\rm v}(S_0)$ is estimated as
\begin{eqnarray}\label{N_sigma}
|N_{\sigma}|
 \lesssim S_{\rm total} \int_{S_{\rm min}}^{S_0}  |n_{\rm v}(S)| \rho dS
=\tilde{S}_{\rm total} \tilde{S}_{\rm min}^{-1/4} \int_{\tilde{S}_{\rm min}}^{\tilde{S}_0} |\tilde{n}_{\rm v}(\tilde{S})| \tilde{\rho} d\tilde{S}={\cal N}(\tilde{S}_0,\tilde{S}_{\rm min})
\end{eqnarray}
with
\begin{eqnarray}\label{vorticity_N}
{\cal N}(\tilde{S}_0,\tilde{S}_{\rm min})=  \left\{
\begin{array}{ll}
\tilde{S}_0^{\frac{3}{4}} \ln\(\frac{\tilde{S}_0}{\tilde{S}_{\rm min}}\)  & \( \tau_{\rm mic}= \frac{5}{4} \) \\
  \tilde{S}_0^{2-\tau_{\rm mic}} - \tilde{S}_0^{\frac{3}{4}}\tilde{S}_{\rm min}^{\frac{5}{4}-\tau_{\rm mic}}  & \( \frac{3}{2}> \tau_{\rm mic} \neq \frac{5}{4} \)
\end{array}
\right. .
\end{eqnarray}
In the case of $\tau_{\rm mic}= \frac{5}{4}$,
the inequality ${\cal N}(\tilde{S}_0,\tilde{S}_{\rm min}) \ll 1$ is satisfied when $\tilde{S}_{\rm min}\ll \tilde{S}_0 e^{-\tilde{S}_{\rm min}^{-\frac{3}{4}}}$.
For the case of $\frac{3}{2}> \tau_{\rm mic} \neq \frac{5}{4}$,
the inequality is safely satisfied with $\tau_{\rm mic} < \frac{5}{4}$.
When $\frac{3}{2} > \tau_{\rm mic} > \frac{5}{4}$,
the inequality is fulfilled for $\tilde{S}_{\rm min} \ll \tilde{S}_0 \ll \tilde{S}_{\rm min}^{\frac{1}{3}}$.

We demand that a statistical behavior should be uniquely determined in a scaling regime.
Since the ${\cal N}$ gives the upper bound of $|N_{\sigma}|$,
we may apply the relation ${\cal N}\ll1$ into the whole range of $\tilde{S}$ in the microscopic regime in the limit $\tilde{S}_{\rm min}\to 0$.
As a result, we may neglect $N_{\sigma}$ in Eq. (\ref{Gamma_nv}) and then one obtains
\begin{eqnarray}\label{nv_approx}
n_{\rm v}(S_0) \approx \frac{\Gamma_0}{\kappa}.
\end{eqnarray}
Accordingly, it has been proved that the vorticity charged along the outer perimeter $L_0$ is computed statistically by neglecting the  contribution from embedded domains, which was assumed empirically in Ref. \cite{2018Takeuchi}.

\section{Summary and discussion}
I evaluated the statistical contribution from internal structure of a circular domain in the microscopic regime of the domain-area distribution in a coarsening development of spontaneous $Z_2$-symmetry breaking in two dimensions.
The area, length and superfluid circulation of the domains may be computed statistically by neglecting the internal structure.
This fact makes it much easier to evaluate the statistical quantities in the microscopic regime
since we may assume that domains in the microscopic regime are independent and they do not have holes of the opposite domain within.

The approximation (\ref{Stotal_approx}) for the domain area is consistent with the observations that
the scaling behavior of the number distribution of domain areas in the microscopic regime is similar to that of hull-enclosed areas in the non-conserved and conserved systems at low temperatures \cite{2007Arenzon}\cite{2008Sicilia}\cite{2009Sicilia},
where the hull-enclosed area refers to the area $S_{\rm total}$ enclosed by the outer perimeter ($L_0$).
The approximation (\ref{nv_approx}) for superfluid circulation supports also the assumption for deriving the dynamic scaling law (\ref{law_nv}) in \Ref{2018Takeuchi}.
For establishing a universal dynamic scaling theory for coarsening dynamics,
I hope the experimental and numerical observations will be done on domain structure in different ordered media from classical systems of binary alloys, binary liquids, and twisted nematic liquid crystal films to quantum systems of two-component and spinor Bose--Einstein condensates, and chiral superfluid $^3$He-A in a slab, and so on.

\begin{acknowledgements}
This work was supported by JSPS KAKENHI Grants No. JP17K05549 and No. JP17H02938. The present research
was also supported in part by the Osaka City University (OCU)
Strategic Research Grant 2018 for young researchers.
\end{acknowledgements}



\end{document}